\documentclass{article}
\usepackage{amsmath}
\usepackage{amssymb}
\usepackage{graphicx}
\usepackage{verbatim,hyperref}
\begin{document}
%\@addtoreset{equation}{section}
%\numberwithin{equation}{section}
%\renewvommand{\theequation}{\thesection.\arabic{equation}}
\newcommand{\Frac}[2]{{\displaystyle \frac{#1}{#2}}}
\def\stackunder#1#2{\mathrel{\mathop{#2}\limits_{#1}}}
\newcommand{\Fig}[2]{%
\begin{center}
\parbox{10cm}{%
\refstepcounter{figure}\includegraphics[width=10cm]{#1} \noindent Figure \thefigure:\quad
#2}\end{center}}
\begin{center}
{\bf\large KINETIC EQUATIONS FOR ULTRARELATIVISTIC PARTICLES\\[2pt] IN
A ROBERTSON-WALKER UNIVERSE AND\\[2pt]  ISOTROPIZATION OF
RELICT RADIATION BY\\[2pt] GRAVITATIONAL INTERACTIONS}\footnote{Original article is
published in \cite{Ast}. In the given version original typing errors are corrected.}\\[8pt]
Yu.G. Ignat'ev and A.A. Popov\\[2pt]
Department of Mathematics, Kazan Teachers Training Institute,
Kazan, U.S.S.R.\\[4pt]
{\footnotesize (Received 31 May, 1989)}
\end{center}

\begin{abstract}
Kinetic equations for ultrarelativistic particles with due account
of gravitational interactions with massive particles in the
Robertson-Walker universe are obtained. On the basis of an exact
solution of the kinetic equations thus obtained, a conclusion is
made as to the high degree of the uniformity of the relict
radiation on scales with are less than $10'$.
\end{abstract}

\section{Introduction} Bisnovatyi-Kogan and Shykhman (1982) have
shown that within the scope of a macro-scopically homogeneous
isotropic cosmological model based on the Newtonian theory of
gravitation we can form kinetic equations with an integral of
gravitational collisions which converge at large distance of
action. The kernel of the collision integral obtained by
Bisnovatyi-Kogan (1982) is the same as that of the Landau
collision integral in Landau (1937). However the 'Coulomb'
logarithm turns out to be finite at any finite cosmological time.
Kandrup (1982) substantiated this result, but the collision
integral differs from the one obtained in Bisnovatyi-Kogan and
Shykhman (1982) in so far it does not vanish on account of
equilibrium distribution. The difference between the kernels of
collision integrals in Bisnovatyi-Kogan and Shykhman (1982) and
Kandrup (1982) seems to be caused by the following. A
Robertson-Walker universe lacks homogeneity in time, which is why
the energy of the particles is not an integral of motion. On the
other hand, in consequence of the long-range character of
gravitational interaction the act of gravitational collision is
protracted. It is precisely the combination of the temporal
non-locality of gravitational interaction and the absence of
temporal homogeneity that leads to the collision integral in
Kandrup (1982). In Zakharov (1984) an integral of Coulomb
collisions is obtained for nonrelativistic charged particles in a
Robertson-Walker universe filled with dust. This integral
coincides with the collision integral obtained by Kandrup, if the
same notations are used, although all calculations in Zahkarov
(1984) are made within the framework of Einstein's theory of
gravitation. It should be noted that the employment of Einstein's
theory of gravitation for the problem of a Coulomb interaction of
non-relativistic particles is unwarranted, since non-relativistic
particles interact by means of Coulomb's field in an identical
manner both in Einstein's and in Newton's treatment, whereas the
Coulomb component of an electromagnetic field in an isotropic
space, being bound up with the law of the conser-  vation of a
charge, does not differ from the traditional component. The are
precisely these facts that account for the agreement between the
results in Zakharov (1984) and Kandrup (1982).

Thus, the derivation of kinetic equations without regard to the
radiation for non-relativistic particles in a Robertson-Walker
universe can always be carried out within the scope of the
Newtonian theory of gravitation and Newtonian mechanics. This
assertion is based on two facts: (1) the adequacy of describing
the effect of a gravitational field on non-relativistic particles
within the scope of Einstein's and Newton's theories of
gravitation, which makes it possible to employ Newtonian
mechanics; (2) the smallness of the path of a non-relativistic
particle as compared to the light horizon, which makes it possible
to employ the Newtonian theory of gravitation.

There exists a fundamentally different condition in the case of
ultrarelativistic particles. A descrition of the gravitational
effect on these particles can only be carried out within the scope
of a relativistic theory of gravitation. Indeed, on the one hand,
the motion of an ultrarelativistic particle is affected by the
components of a three-dimensional metric, which is why Newtonian
mechanics cannot be employed to describe motion. On the other
hand, an ultrarelativistic particle runs a distance comparable to
the light horizon, which makes the Newtonian description of a
gravitational field unacceptable. However, in the case of
electromagnetic interaction the question of forming kinetic
equations for ultrarelativistic particles in a Robertson-Walker
universe is easily solved; in consequence of the conformal
invariance of field equations and motion equations, these
equations will be not different from those in a plane space (see
Ignat'ev, 1982).

This brings to the fore the basic problem of formulating kinetic
equations for ultrarelativistic particles in a Robertson-Walker
universe with allowance for their gravitational interaction with
non-relativistic ones. Suitable techniques for forming these
equations are provided by Ignat'ev (1978, 1983). In the present
paper a kinetic equation for ultrarelativistic particles is
derived, with due account of their gravitational interaction with
massless ones. This is carried out on the basis of averaging a
collisionless kinetic equation over the local fluctuations of the
Robertson-Walker metric caused by the gravitational fields of
massive point particles. The solution of Einstein's equations
linearized about a Robertson-Walker solution shows that the local
fields of point masses formed by a redistribution of
Robertson-Walker matter, always remain within the sound horizon.
In this connection the problem of the convergence of the collision
integral at large distance of action becomes irrelevant. The
kinetic equation thus obtained has a collision term whose
structure is the same as that of the Belyaev-Budker collision
integral (cf. Belayev and Budker, 1956), if one of the particles
is taken to be non-relativistic in the latter. A solution of the
kinetic equation thus obtained is presented. The solution
describes the process of the isotropization of a homogeneous, but
anisotropic distribution of massless relict particles. The
estimates show that angular harmonics with a scale less than 10
angular minutes are strongly damped. All the notation, unless
otherwise specified in the text, are according to those in
%Ignat'ev (1978, 1981, 1982, 1983), Belayev and Budker (1956), or
Landau and Lifshitz (1972).

\section{Massive Particles in a Robertson-Walker Universe}
The gravitational field of a massive point particle in a
Robertson-Walker universe is basically obtainable with the aid of
the well-known Lifshitz solutions (see Landau and Lifshitz, 1972).
However, the formulation of initial and boundary conditions in a
problem with spherical symmetry in terms of plane waves loses its
physical lucidity. Besides, the synchronous frame of reference by
Landau and Lifshitz (1972) is unsuitable for our purpose. We shall
write the energy-momentum tensor of a massive point particle
(Ignat'ev, 1983) as ($\hbar=c=G=1$):
\begin{equation}\label{eq2.1}
\delta T^{ik}=m\int u^i u^k {\rm D}(x|x')dS
\end{equation}
where $D(x/x')$ is the invariant four-dimensional function of
Dirac: i.e.
\[\int D(x|x')d\Omega = 1\]
($d\Omega = \sqrt{-g}d^4 x$); integration in \eqref{eq2.1} is carried out
along the entire world line of the particle $x' = x'(S)$. As will
be seen from the following, in a non-empty space the mass of a
particle cannot remain constant; therefore, $m = m(x)$ is a scalar
function. The motion equations of a point particle with a variable
mass is obtainable from an invariant Hamiltonian (Ignat'ev, 1982)
\[H_m(x,P)= \frac{1}{m}g^{ik}P_iP_k-m=0,\]
from which we find using standard procedure (as in Ignat'ev, 1983)
\begin{equation}\label{eq2.2}
\frac{du^i}{dS}=(\ln m)_{,k}(g^{ik}-u^iu^k).
\end{equation}
If we assume the spherical symmetry of the problem we shall write
the space-time metric in isotropic coordinates
\begin{equation}\label{eq2.3}
dS^2 = e^\nu d\nu^2 - e^\lambda [dr^2 + r^2 (d\theta^2 +
\sin^2\theta d\varphi^2)] ,
\end{equation}
where $\nu = \nu(r, \eta)$, $\lambda=\lambda = (r, \eta)$. Then
the world line of the particle is the time line $r = 0, m =
m(\eta)$ being an arbitrary function in consequence of \eqref{eq2.2}.
Integrating \eqref{eq2.1} we get
\[\delta T^i_k=\delta^i_4\delta^4_k m e^{3/2\lambda}\delta(\mathbf{r}) \]
where $\delta(\mathbf{r})$ is Dirac's three-dimensional $\delta$ -
function in plane space. We shall represent the total
energy-momentum tensor $T^{ik}$ as
\[T^{ik} =T^{ik}_f + \delta T^{ik},\]
where $T^{ik}_f$ is the energy-momentum tensor of a fluid. In
consequence of the Einstein equations, \eqref{eq2.1} and \eqref{eq2.2} the law of
conservation is satisfied by
\[T^{ik}_{f~,k}=-\nabla^i\int m\mbox{D}(x|x')dS.\]
In particular, it is, in our case, of the form
\[T^{4k}_{f~,k}= -\dot{m}e^{-(\mu+\lambda)}\delta(\mathbf{r}).\]

Einstein's non-trivial equations for the metric \eqref{eq2.3} take the
form (see Landau and Lifshitz, 1972)
\[
\frac{1}{2}e^{-\lambda}\left[\Frac{\lambda^{'2}}{2} + \lambda'
\nu\,' + \Frac{2}{r}(\lambda' + \nu\,')\right] -
e^{-\nu}\left(\ddot{\lambda} - \frac{1}{2}\dot{\lambda}\dot{\nu} +
\frac{3}{4}\dot{\lambda}^2 \right)=8\pi[p + v^2(\varepsilon + p)];
\]
\[
\frac{1}{4}e^{-\lambda}\left[2(\lambda''+\nu'')+\nu'\
^2+\frac{2}{r}(\lambda'+\nu')\right] - e^{-\nu}\left(\ddot{\lambda}
- \frac{1}{2}\dot{\lambda}\dot{\nu} + \frac{3}{4}\dot{\lambda}^2
\right)=8\pi p;
\]
\[ -e^{-\lambda}\left(\lambda'' + \Frac{\lambda^{'2}}{4} +
\frac{2}{r}\lambda' \right) + \Frac{3}{4}e^{-\nu}\dot{\lambda}^2 =
8\pi \left[me^{-\frac{3}{2}\lambda}\delta(\vec{r}) + \varepsilon +
v^2(\varepsilon + P)\right]; \]
\[\frac{1}{2}e^{-\lambda}(2\dot{\lambda}' - \nu\,' \dot{\lambda}) =
8\pi(\varepsilon + p)e^{\frac{1}{2}(\nu + \lambda)}v\sqrt{1 +
v^2},\]
where $v=u^re^{\lambda/2}$ is the reference projection of the radial
velocity of a fluid. If we subtract the second equation from the
first, we get
\[\frac{1}{2}e^{-\lambda}\left[\frac{\lambda'\ ^2}{2}+\lambda'\nu' -\frac{\nu'\ ^2}{2}
+\frac{1}{r}(\lambda'+\nu')-(\lambda''+\nu'')\right]=8\pi(\varepsilon+p)v^2.\]
We shall take $m, v, \lambda', \nu'$ to be first-order
infinitesimals. Then in a linear approximation the last equation is
easily integrable, yielding
\[\lambda+\nu=C_1(\eta)r^2+C_2(\eta).\]
By making a direct substitution we can very that $C_1= 0$.
Admissible transformations of coordinates which preserve the form of
the metric \eqref{eq2.3} are
\[\eta=\eta(\tilde{\eta});\quad r=k\tilde{r};\quad (k=\mbox{Const}). \]
In consequence, we can add to $\nu$ an arbitrary time function, and
to $\lambda$ an arbitrary constant. Let us select this function so
that
\begin{equation}\label{eq2.4}\lambda=\ln a^2+\xi(r,\eta),\quad \nu=\ln
a^2-\xi(r,\eta),\end{equation}
where $a = a(\eta)$,  $\xi\ll 1$. Linearizing Einstein's equations
with respect to the smallness of $m, v,\xi$, we obtain the system
\begin{equation}\label{eq2.5}
\frac{1}{a^2}\left(\Frac{2\ddot{a}}{a}-\Frac{\dot{a}^2}{a^2}\right)=-8\pi
p_0;\quad 3\Frac{\dot{a}^2}{a^4}=8\pi\varepsilon_0;
\end{equation}
\begin{equation}\label{eq2.6}
\ddot{\xi}+3\dot{\xi}\frac{\dot{a}}{a}+\left(\Frac{2\ddot{a}}{a}-\Frac{\dot{a}^2}{a^2}\right)\xi=-8\pi
a^2\frac{dp}{d\varepsilon} \delta\varepsilon;
\end{equation}
\begin{equation}\label{eq2.7}
-\frac{1}{a^2r^2}\frac{\partial}{\partial
r}(r^2\xi')+3\frac{\dot{a}}{a^4}\frac{\partial}{\partial
\eta}(a\xi)=8\pi \frac{m}{a^3}+8\pi\delta\varepsilon;
\end{equation}
\begin{equation}\label{eq2.8}
v=\frac{1}{8\pi(\varepsilon_0+p_0)}\frac{\partial}{\partial\eta}(a\xi'),
\end{equation}
where $p_0=p_0(\eta)$, $\varepsilon_0 =\varepsilon_0(\eta)$, and we
have put $\delta p = (dp/d\varepsilon \delta\varepsilon$. Equations
(2.5) describe the evolution of a flat-space Robertson-Walker
universe (cf. Landau and Lifshitz, 1972); Equation \eqref{eq2.8} determines
the radial velocity of a fluid. To solve the singular equations
\eqref{eq2.6} and \eqref{eq2.7}, let us put
\begin{equation}\label{eq2.9}
\xi=\frac{2}{ra}(m-\Psi),
\end{equation}
and $m = m(\eta)$,
\begin{equation}\label{eq2.10}
\lim_{r\to 0} \Psi(r,\eta)/r < + \infty.
\end{equation}
If we substitute \eqref{eq2.9} into \eqref{eq2.7} and taking account of \eqref{eq2.10} we
get
\begin{equation}\label{eq2.11}
4\pi
a^3\delta\varepsilon=\frac{1}{r}\Psi''+\frac{3\dot{a}}{ar}\frac{\partial}{\partial\eta}(m-\Psi).
\end{equation}
By use of (2.11) in the right-hand side of Equation 2.6), we get
closed equations for $m(\eta)$ and $\Psi(r,\eta)$:
\begin{equation}\label{eq2.12}
\ddot{m}+\frac{\dot{a}}{a}\dot{m}\left(1+3\frac{dp}{d\varepsilon}\right)+
m\left(\Frac{2\ddot{a}}{a}-\Frac{\dot{a}^2}{a^2}\right)=0;
\end{equation}
\begin{equation}\label{eq2.13}
\ddot{\Psi}+\frac{\dot{a}}{a}\dot{\Psi}\left(1+3\frac{dp}{d\varepsilon}\right)+
\Psi\left(\Frac{2\ddot{a}}{a}-\Frac{\dot{a}^2}{a^2}\right)=\Psi''\frac{dp}{d\varepsilon}.
\end{equation}
At the non-relativistic stage $p_0 = 0$ and, according to \eqref{eq2.5},
$a\sim \eta^2$. Then \eqref{eq2.12} and \eqref{eq2.13} are easily integrable:
yielding
\begin{equation}\label{eq2.14}
m=\sigma\left(\frac{\eta_0}{\eta}\right)^3+\mu\left(\frac{\eta}{\eta_0}\right)^2;\quad
\Psi=W(r)\left(\frac{\eta_0}{\eta}\right)^3+V(r)\left(\frac{\eta}{\eta_0}\right)^2,
\end{equation}
where $\sigma, \mu$ are arbitrary constants and $W(r), V(r)$ are
arbitrary functions. Thus, the mass of a point particle in a
medium is not conserved, which is the consequence of its
gravitational interaction with the fluid.

To determine the functions $W(r), V(r)$ it is necessary to
consider the evolution of a metric at the ultrarelativistic stage
of expansion ($\varepsilon_0 = 3p_0$). In this case Equations
\eqref{eq2.12} and \eqref{eq2.13} take the form
\[ \ddot{m}+\frac{2}{\eta}\dot{m}-\frac{2}{\eta^2}m=0;\quad
\ddot{\Psi}+\frac{2}{\eta}\dot{\Psi}-\frac{2}{\eta^2}\Psi=\frac{1}{3}\Psi''. \]
The first one is easily integrated, and yields
\begin{equation}\label{eq2.15}
m=\sigma'\left(\frac{\eta_0}{\eta}\right)^2+\mu'\frac{\eta}{\eta_0},
\end{equation}
where $\sigma', \mu'$ are arbitrary constants; to solve the second
equation we shall make a substitution $\eta=\sqrt{3}\tau$ and
\begin{equation}\label{eq2.16}
\Psi=\frac{\partial}{\partial\tau}\frac{1}{\tau}\Phi(r,\tau).
\end{equation}
On substituting \eqref{eq2.16} into the second equation and changing the
order of differentiation we shall reduce it to the form
\begin{equation}\label{eq2.17}
\frac{\partial}{\partial\tau}\frac{1}{\tau}(\Phi_{\tau\tau}-\Phi_{rr}),
\end{equation}
whence
\begin{equation}\label{eq2.18}
\Phi(r,\tau)=\Phi_+(r+\tau)+\Phi_-(r,\tau).
\end{equation}
To the solution obtained we can add $\tau F(r)$, where $F(r)$ is
an arbitrary function. However, this addition, according to
\eqref{eq2.16}, does not alter the value of $\Psi$. A particular solution
to \eqref{eq2.18} is $\Phi=0$. Then, according to \eqref{eq2.9}, we shall obtain a
Newtonian potential caused by a point particle of variable mass
$m(\eta)$ determined with the aid of \eqref{eq2.15}. However, this
solution is physically unacceptable for the description of the
gravitational field of a point particle which arose as a result of
fluctuation at the moment of time $t = 0$, since it is
inconsistent with the principle of causality. Indeed, the
Newtonian potential referred to make it possible to obtain
information about a particle beyond its light horizon and even to
determine the mass of the particle. Therefore, boundary conditions
should be formulated in such a way that at least beyond the light
horizon the potential $\xi$ should vanish together with its
derivatives. Such boundary conditions are in according with the
'birth' of a particle as a result of the redistribution of
Robertson-Walker matter. In fact, however, the horizon of the
potential is not the light cone but the sound cone, since the
change in the potential is bound up with the redistribution of
matter, and the latter proceeds precisely at the speed of sound.
Mathematically this is related to the characteristics of Equation
(2.17) and it is precisely at the sound horizon $r = \tau =
r/\sqrt{3}$ that boundary conditions take the simplest form.

For the mass of a particle to remain limited at $t\to 0$ it is
necessary to put $\sigma'= 0$ in the solution \eqref{eq2.15}. Boundary
conditions at the light horizon which satisfy the principle of
causality are of the form
\[ \xi(r,\tau)\bigr|_{r=\tau}=0;
\quad \xi'(r,\tau)\bigr|_{r=\tau}=0;\quad v(r,\tau)\bigr|_{r=\tau}=0. \]
The latter condition is, however, automatically satisfied in
consequence of the first two conditions. If we substitute $\xi$
here from \eqref{eq2.9} we shall reduce the first two conditions to the
form
\begin{equation}\label{eq2.19}
\Psi(\tau,\tau)=m(\tau)=\mu\frac{\tau}{\tau_0};\quad
\left.\Psi(r,\tau)\right|_{r=\tau}=0.
\end{equation}

According to the definitions of \eqref{eq2.17} and \eqref{eq2.18} these conditions
are equivalent to
\begin{equation}\label{eq2.20}
\Phi_+(2\tau)+\Phi_-(0)=2\frac{\mu'(\tau)}{\tau_0}\tau^3+A\tau;
\end{equation}
\begin{equation}\label{eq2.21}
\Phi'_+(r+\tau)\bigr|_{r=\tau}-\Phi'_-(r-\tau)\bigr|_{r=\tau}=B\tau,
\end{equation}
where $A$ and $B$ are arbitrary constants. Besides, due to \eqref{eq2.10}
another condition is to be satisfied: namely
\begin{equation}\label{eq2.22}
\Phi_+(\tau)+\Phi_-(\tau)=0.
\end{equation}
From \eqref{eq2.20} and \eqref{eq2.22} we find that
\[\Phi_+(x)=\frac{1}{4}\frac{\mu'}{\tau_0}x^3 +\frac{A}{2}x+\Phi_+(0),\]
\[\Phi_-(x)=-\frac{1}{4}\frac{\mu'}{\tau_0}x^3 -\frac{A}{2}x-\Phi_+(0). \]
Differentiating these relations and substituting the results into
\eqref{eq2.21}, we get $SA=0$, $B=-3/\mu'/\tau_0$. If we substitute
the obtained value of $\Phi(r,\tau)$ into \eqref{eq2.16}, we finally
get\footnote{A solution to \eqref{eq2.23} is also obtainable as an
auto-model one, putting $\Psi = \tau\overline{\Psi}(r/\tau)$.}
\begin{equation}\label{eq2.23}
\Psi(r,\tau)=\left\{\begin{array}{ll} {\displaystyle
\frac{1}{2}\mu'\frac{r}{\tau_0}\left(3-\frac{r^2}{\tau^2}
\right)};& r\leq \tau;\\[10pt]
{\displaystyle-\mu'\frac{\tau}{\tau_0}}, & r>\tau.
\end{array}
\right.
\end{equation}
Let $\eta_0$ be a moment of 'time' when the ultrarelativistic
stage is replaced by the non-rela\-tivistic. Suppressing in the
solution of \eqref{eq2.14} terms which correspond to the dispersion of the
mass and joining this solution to \eqref{eq2.23} at the moment $\eta_0$,
we get an expression for the potential $\xi(r,\tau)$ at the
non-relativistic stage of expansion
\begin{equation}\label{eq2.24}
\xi(r,\tau)=\left\{\begin{array}{ll} {\displaystyle
\frac{2\mu}{r}\left[1-\frac{r}{2\tau_0}\left(3-\frac{r^2}{\tau_0^2}
\right)\right]};& r\leq \tau;\\[10pt]
0, & r>\tau.
\end{array}
\right.
\end{equation}

The potential obtained is time-independent. Therefore, if the
distance $\Delta r$ from a certain observer, synchronous in a
Robertson-Walker metric, to a massive particle is more than the
value $\eta_0/\sqrt{3}$, the observer will never be affected by
the gravitation of a particle. Thus, in a Robertson-Walker
universe the effective range of the gravitational forces of point
particles turns out to be finite. The ratio between this range
$l_g = a(\eta)\eta_0/\sqrt{3}$ and the distance to the light
horizon $l_c$ at the non-relativistic stage of expansion is less
than unity and decreases as time goes on
\begin{equation}\label{eq2.25}
\frac{l_g}{l_c}=\frac{1}{\sqrt{3}}\left(\frac{t_0}{t}\right)^{1/3}.
\end{equation}
At $t_0 \sim 10^{13}$c, $t\sim 10^{18}$c, $l_g/l_c \sim 10^{- 2}$.
It is of interest to note that the dimension of gravitationally
bound regions $l_g$ in the present epoch turns out to be of the
order of 100 Mpc, which at the medium density $\varrho=10^{-29}$
$\mbox{g}\cdot \mbox{cm}^{-3}$  Mpc$^{-1}$ indicates a mass of the
order of $10^{18} M_\odot$. It is also worth of noting that the
mass $m \sim m_{pl}$ per a Planck moment of time increases to
values of the order of $10^{26}$g in the present epoch.

If we substitute Equation \eqref{eq2.24} into \eqref{eq2.11}, we shall obtain an
expression for the perturbation of the energy density caused by a
massive particle, of the form
\begin{equation}\label{eq2.26}
{\displaystyle
\frac{\delta\varepsilon}{\varepsilon_0}=-\frac{-m(\eta)}{\frac{4}{3}\pi
r^3_0a^3 \varepsilon_0}+\xi(r)\equiv
\frac{m(\eta)\sqrt{3}}{2t_0}+\xi(r), }\end{equation}
from which follows an obvious fact that the perturbation of the
energy density remains small as long as the mass of pa particle is
small as compared to the mass of the gravitationally bound region:
\begin{equation}\label{eq2.27}
m(\eta)\ll \frac{4}{3}\pi r^3_0
a^3\varepsilon_0(\eta)=\frac{2t_0}{\sqrt{3}}\equiv \overline{m}.
\end{equation}

At $t_0 = 10^{13}$ s we have $m = 10^{18}M_\odot$. According to
(2.26), at the boundary of the gravitationally bound region
($r=r0$) there arises an abrupt change in density
$\delta\varepsilon(r0,\eta)/\varepsilon_0(\eta) = -
m(\eta)/\overline{m}$, which vanishes at $\eta\to 0$ and increases
as time goes on. At the end the entire matter from within the
sphere with the radius $r_0$ accretes by a massive
particle\footnote{If this is not impeded by the forces of pressure
or a large angular momentum of the dust.}, and the finite
gravitational field in an empty sphere will be described by a
Schwarzschild metric with the gravitational mass $\overline{m}$.
It is not difficult to obtain this finite metric by joining a
Schwarzschild metric in an isotropic coordinate system to a
Robertson-Walker metric (see, for instance, Landau and Lifshitz,
1972) on a sphere with the 'radius' $r_0 = \mbox{const}$. A smooth
joining is possible precisely at $r_g = 2\overline{m}$, where
$\overline{m}$ is described by Equation \eqref{eq2.27}.

\section{Local Fluctuations of the Metric and Their Averages}
Let there be in a Robertson-Walker universe not one, but several
massive identical particles\footnote{If the masses of the
particles are different, it is also necessary to average out the
formulae obtained, with respect to the distribution of the
masses.} with the coordinates $\mathbf{r}_a = \{x_a, y_a,z_a\}$.
Then the total space-time metric can be approximately written in a
form (linear approximation with respect to $m$):
\begin{equation}\label{eq3.1}
dS^2=(g_{ik}+h_{ik})dx^idx^k,
\end{equation}
where $g_{ik}$ is a Robertson-Walker metric,
\begin{equation}\label{eq3.2}
h_{ik}=-a^2\delta_{ik}\sum\limits_{a}\xi_a(|\mathbf{r}-\mathbf{r}_a|).
\end{equation}
Indeed, a contribution to the metric caused by the interaction of
particles is of the order $m^2$; non-diagonal components of the
metric tensor $g_{\alpha 4} \sim \xi v_\alpha  \sim m^2$ are also
of the same order. Introduce a field of observers, macroscopic in
the metric of \eqref{eq3.1}, whose coordinate grid is on massive
particles. Such observers are geodesic with respect to the
Robertson-Walker metric. Let the coordinates of massive particles
$\mathbf{r}_a$ assume random equiprobable values throughout the
entire three-dimensional space, with no correlation between the
positions of these particles. Let $N = \mbox{Const}$, be the
number of massive particles per Volume' $V = 4/3\pi r^3_0$. Let us
introduce an operation of averaging a certain field value
$\varphi(x|r_l,r_2,\ldots)$, which is a function of the positions
of the massive particles, on the scale of macroscopic observers:
i.e.,
\begin{equation}\label{eq3.3}
\langle\varphi(x)\rangle=\prod\limits_{a}\frac{N}{V_a}\int
d^3\mathbf{r}_a \varphi(x|x_l,x_2,\ldots)
\end{equation}
\newcommand{\average}[1]{\langle #1\rangle}
where integration is carried out within the spheres with the
radius $r_0$ with centres at the point; $V_a = V$.\footnote{In the
formula \eqref{eq3.3} integration could be carried out over the entire
space; however, this would yield nothing new, since according to
\eqref{eq2.24} outside the sphere with the radius correlations vanish.}
Then the average of the local fluctuations of the Robertson-Walker
metric of \eqref{eq3.2}, according to \eqref{eq2.24} and \eqref{eq3.3}, is
\begin{equation}\label{eq3.4}
\average{h_{ik}}=-a^2\delta_{ik}\frac{3\mu
N}{5r_0}=a^2(\eta)\mbox{Const}.
\end{equation}
If we follow the procedure of Ignat'ev (1978), we shall
renormalize the macroscopic Robertson-Walker metric and the local
fluctuations $h_{ik}$ so that the average of the latter should be
equal to zero: i.e.,
\begin{equation}\label{eq3.5}
g_{ik}\to g_{ik}+\average{h_{ik}};\quad h_{ik}\to
h_{ik}-\average{h_{ik}}; \quad \xi_a\to \xi_a-\frac{3\mu}{5r_0}.
\end{equation}
Due to \eqref{eq3.4} a renormaliztion of the macroscopic metric $g_{ik}$
reduces to multiplying $g_{44}$ and $g_{\alpha\beta}$ by constant
numbers. Carrying out a further admissible infinitesimal scale
transformation of the coordinates $\eta$ and $r$ with a constant
scale coefficient,we shall restore the former value of the
Robertson-Walker metric
\begin{equation}\label{eq3.6}
\average{dS^2} = a^2(\eta) (d\eta^2 - dx^2 - dy^2 - dz^2).
\end{equation}
By use of the value of the function of $\xi(r)$ redefined
according to \eqref{eq3.5} we shall evaluate the change in the total mass
$\delta M(r0, \eta)$ within the radius $r_0$ caused by a massive
particle. For this purpose, we shall transform the expression for
the total mass $M(r)$ in the Schwarzschild coordinate system
(Landau and Lifshitz, 1972), linearize it with respect to $\xi$
and employ the formula \eqref{eq2.24}. As a result we get
\[\delta M(r_0,\eta)=4\pi\frac{3\dot{a}}{a}\frac{d}{d\eta}
\int\limits_{0}^{r_0}\xi r^2dr=0  \]
Thus, the solution we have obtained actually describes the
gravitational field of a particle formed by a redistribution of
Robertson-Walker matter.

If we calculate $\average{\xi^2(\mathbf{r})}$ according to \eqref{eq3.3}
and \eqref{eq3.5}, we get
\begin{equation}\label{eq3.7}
\average{\xi^2(\mathbf{r})}=\sum\limits_a
\average{\xi_a^2(\mathbf{r})}=\frac{108}{175}N\left(
\frac{2\mu}{r_0}\right)^2.
\end{equation}
For the local and medium perturbations of the metric to be small,
two conditions are to be satisfied: namely,
\begin{equation}\label{eq3.8}
\bigl|r_a -r\bigr| \gg 2\mu;
\end{equation}
\begin{equation}\label{eq3.9}
N\left(\frac{2\mu}{r_0}\right)^2  \ll 1.
\end{equation}
Since several massive particles can now find themselves within the
sphere with the radius $r_0$, the condition of \eqref{eq2.27} must be
replaced by
\begin{equation}\label{eq3.10}
nm(\eta)\equiv N\mu \left(\frac{t}{t_0} \right)^{2/3}\ll
\overline{m}.
\end{equation}
By use of the properties of isotropy and homogeneity we can show
the validity of the equalities
\begin{equation}\label{eq3.11}
\begin{array}{ll}
{\displaystyle
\average{h_{ik}h_{lm}}=\delta_{ik}\delta_{lm}a^4\average{\xi^2}};
&
\\[10pt]
{\displaystyle{\average{\partial_j(h_{ik})h_{lm}}}=0};&
{\displaystyle{\average{\partial_4(h_{ik})\partial_j( h_{lm})}}=0}.\\
\end{array}
\end{equation}
Averages of the type $\average{\partial_\alpha h_{ik}
\partial_\beta h_{lm}}$ diverge as $r^{-1}$, the divergence of these
values is connected with the ordinary divergence of the particle's
energy. To calculate these values we shall note that, due to the
isotropy of space,
\newcommand{\averxi}[2]{\average{(\partial_{#1}\xi)(\partial_{#2}\xi)}}
\[
\averxi{\alpha}{\beta}=\frac{1}{3}\delta_{\alpha\beta}\averxi{\gamma}{\gamma}=\frac{1}{3}\delta_{\alpha\beta}
[\average{\partial_\gamma (\xi\partial_\gamma \xi)} -
\average{\xi\partial_{\gamma\gamma}\xi}].\]

Thus,
\begin{equation}\label{eq3.12}
\averxi{\alpha}{\beta}=-\frac{1}{3}\average{\xi\Delta\xi}.
\end{equation}
By use of Equation \eqref{eq2.7} and the explicit form of the function
$\Delta\xi$ in \eqref{eq2.9}, \eqref{eq2.24} to calculate $\xi(r)$, we find
\[ -\Delta\xi_a=8\pi\mu\delta(\mathbf{r}-\mathbf{r}_a)-\frac{6\mu}{r_0^3}.  \]
As a result we get
\[ -\average{\xi\partial_{\gamma\gamma}\xi}=\frac{8\pi\mu}{V}\int\delta(\mathbf{r})\xi_a ( \mathbf{r}) dV. \]
This integral diverges as $r^{-1}$, which is caused by the
divergence of the total energy of the particle. To calculate the
integral we shall employ the procedure of renormalizing the mass,
treating $\delta(r)/r$ as $-\delta'(r)$. Then, taking into account
\eqref{eq3.12}, we get eventually
\[-\average{\xi\Delta\xi} =\frac{18\pi N}{r^2_0}\left(\frac{2\mu}{r_0}\right)^2; \]
\[\averxi{\alpha}{\beta}=\frac{6\pi N}{r^2_0}\left(\frac{2\mu}{r_0}\right)^2 \delta_{\alpha\beta}.\]
If we calculate the averages of corrections to the Einstein tensor
caused by the fluctuations of the metric we shall find the
corrections to the energy-momentum tensor of Robertson-Walker dust
caused by the energy of the local gravitational fields:
\[ \delta T^g_{ij}= -\frac{1}{8\pi}\average{\delta G_{ij}} =\frac{g_{ij}}{8\pi a^2}
\average{\xi\Delta
\xi}=-\frac{gN}{4r^2_0}\left(\frac{2\mu}{r_0}\right)^2
\frac{g_{ij}}{a^2}. \]
Thus, we shall obtain corrections to the energy density
$\delta\varepsilon_g$ and pressure $\delta p_g$ of the
form\footnote{In original article \cite{Ast} because of negligence
the error in a sign on density of energy $\delta\varepsilon_g$ is
admitted. However, expression for $\delta T^g_{ij}$ is resulted
the correct.}
\begin{equation}\label{eq3.13}
\delta\varepsilon_g=-\delta p_g =
-\frac{gN}{4(ar_0)^2}\left(\frac{2\mu}{r_0}\right)^2\Rightarrow
\delta\varepsilon_g+\delta p=0.
\end{equation}
The allowance for the local fluctuations of the gravitational
field in the Einstein equations is equivalent to adding to the
Robertson-Walker dust a fluid with the equation of state
$\varepsilon=-p$. Note that the ratio
$\delta\varepsilon_g/\varepsilon_0$ increases in proportion to
$\eta^2$, but remains small when the conditions of \eqref{eq3.8}--\eqref{eq3.10}
are satisfied.

\section{Derivation of the Kinetic Equation}
Let $\bar{f}(x,\bar{P}|x_1,x_2,\ldots) = \bar{f}(x,\bar{P})$ be a
macroscopic function of the distribution of massless particles;
then the total number of particles recorded by the observers,
associated with the velocity field $u^i$ on hypersurface $\Sigma$
orthogonal to this field (cf. Ignat'ev, 1983) is
\begin{equation}\label{eq4.1}
\bar{L}=\iiint\limits_\Sigma d\Sigma\int d^4 \bar{P}u^i
\bar{P}_i\delta(\bar{H})\bar{f}(x,\bar{P}),
\end{equation}
where $d^4 \bar{P} = d\bar{P}_1 d\bar{P}_2 d\bar{P}_3
d\bar{P}_4/\sqrt{-g}$; $H(x, P)$ is the invariant Hamiltonian of
massless particles. We have used a bar to note the fact that all
the values determined in a Robertson-Walker universe perturbed by
local fluctuations. In accordance with our procedure we shall
select as $u^i$ a field of observers which is geodesic in a
macroscopic Robertson-Walker universe, then
$u^i=\delta^i_4/\sqrt{g_{44}}$ . Thus, taking into account \eqref{eq3.1},
\eqref{eq3.2}, \eqref{eq3.4}, and \eqref{eq3.5}, we get from \eqref{eq4.1}
\begin{equation}\label{eq4.2}
\bar{L}=\iiint\frac{d^3x}{\bar{g}_{44}} \int d^4 \bar{P}
\bar{P}_4\delta(\bar{H})\bar{f}(x,\bar{P}).
\end{equation}
We shall take into account the fact that the Hamiltonian of
massless particles is of the form
\begin{equation}\label{eq4.3}
\bar{H}(x,\bar{P})=\frac{1}{2} g^{ik}P_i
P_k=\frac{1}{2a^2}\left(\frac{\bar{P}_4^2}{1-\xi}-\frac{\bar{P}^2}{1+\xi}
\right),
\end{equation}
where $\bar{P}^2=\bar{P}^2_1+\bar{P}^2_2+\bar{P}^2_3$. Transform
the formulae \eqref{eq4.2} and \eqref{eq4.3} to the new variables $x^i, P_i$:
\begin{equation}\label{eq4.4}
P_\alpha=\bar{P}_\alpha; \quad
P_4=\bar{P}_4\sqrt{\frac{1+\xi}{1-\xi}};
\end{equation}
then
\begin{equation}\label{eq4.5}
\bar{H}(x,\bar{P})=\frac{1}{1+\xi} H(x,P),
\end{equation}
where $H(x,P)$ is the Hamiltonian of massless particles in an
unperturbed Robertson-Walker universe (3.6). Modifying the formula
\eqref{eq4.2} taking into account the properties of $\delta(\bar{H})$, we
get
\begin{equation}\label{eq4.6}
\bar{L}=\iiint\frac{d^3x}{g^{44}} \int d^4 P\
P_4\delta(H)\bar{f}(x,\bar{P}),
\end{equation}
which is precisely the same expression as the one in terms of an
unperturbed metric \eqref{eq3.6}. It should be noted that in \eqref{eq4.6} the
fluctuation of the metric is present only in the function of the
distribution of $f(x,P)$. Averaging \eqref{eq4.6} according to \eqref{eq3.3} we
find that
\begin{equation}\label{eq4.7}
\average{L}\equiv L=\iiint\limits_\Sigma d\Sigma \int d^4 P\
u^iP_i\delta(H)f(x,P),
\end{equation}
where
\begin{equation}\label{eq4.8}
f(x,P)\equiv \average{\bar{f}(x,\bar{P})}.
\end{equation}
Thus, the average number of massless particles is evaluated with
the aid of an ordinary formula, in a macroscopic Robertson-Walker
universe, with respect to the distribution function averaged over
the local fluctuations of the metric. This is the advantage of the
frame of reference that we have selected. If we had selected, for
instance, a macroscopi\-cally synchronous frame of reference, then
instead of the simple relationship (4.8) we would have had a
complex relationship which would include the derivatives of the
distribution function f(x, P) and the correlations of this
function with the local fields. This is caused by the fact that a
macroscopic ally local observer is subjected to the effect of the
local gravitational fields, as a result of which the scales of
clocks and rods on the micro -- and macro -- levels are different.

Now we shall proceed to average the microscopic collisionless
kinetic equation (as in Ignat'ev, 1981)
\begin{equation}\label{eq4.9}
\delta(H)\left[{\displaystyle \frac{\partial \bar{H}}{\partial
\bar{P}_i}\frac{\partial \bar{f}}{\partial x^i}%
- \frac{\partial \bar{H}}{\partial x^i}%
\frac{\partial \bar{f}}{\partial \bar{P}_i}}\right]=0
\end{equation}
If we carry out in \eqref{eq4.9}) a preliminary transformation to the new
variables $x^i,P_i$ according to (4.4), we get
\begin{eqnarray}\label{eq4.10}
\delta(H)\left\{\frac{\partial H}{\partial P_\alpha}\frac{\partial
f}{\partial x^\alpha}+\sqrt{\frac{1+\xi}{1-\xi}}\frac{\partial
H}{\partial P_4}\frac{\partial f}{\partial x^4} \right.\hspace{5cm}\nonumber\\%
\hspace{3cm}+\frac{1}{2}\left(\partial_\alpha\ln\frac{1+\xi}{1-\xi}\right)\left.\left[%
\frac{\partial H}{\partial P_\alpha}P_4\frac{\partial f}{\partial
P_4} -%
\frac{\partial H}{\partial P_4}P_4\frac{\partial f}{\partial
P_\alpha}
\right]\right\}.
\end{eqnarray}
In Equation \eqref{eq4.10} all the momentum variables $P_\alpha$ and $P_4$
are treated as independent. Expressing $P_4$ in terms of
$P_\alpha$ with the aid of the mass shell equation $H(x, P) = 0$,
we shall reduce (4.10) to a simpler form
\begin{equation}\label{eq4.11}
P^\alpha\frac{\partial f}{\partial
x^\alpha}+\sqrt{\frac{1+\xi}{1-\xi}}P_4\frac{\partial f}{\partial
\eta}-\frac{1}{2}\left(\partial_\alpha\ln\frac{1+\xi}{1-\xi}\right)P_4P^4\frac{\partial
f}{\partial x^\alpha}=0
\end{equation}
($P^\alpha = -P_\alpha/a^2; P^4 = P_4/a^2$). Our task is to obtain
a kinetic equation for the macroscopic distribution function
$f(x,P)$ to within an accuracy of terms quadratic with respect to
the local fluctuations of the metric. Therefore, we shall expand
\eqref{eq4.11} into a series with respect to the smallness of $\xi$
confining ourselves to the second-order terms:
\begin{equation}\label{eq4.12}
P^\alpha\frac{\partial \bar{f}}{\partial
x^\alpha}+\left(1+\xi+\frac{\xi^2}{2} \right) P_4\frac{\partial
\bar{f}}{\partial \eta}-(\partial_\alpha\xi)P_4P^4\frac{\partial
\bar{f}}{\partial x^\alpha}=0.
\end{equation}
Taking into account the fact that correlations between the
positions of massive particles are absent, we shall represent the
macroscopic distribution function f(x, P) in the form
\[\bar{f}(x,P)=f(x,P) + g(x,P),\]
where
\begin{equation}\label{eq4.13}
g(x,P|x_l,x2,\ldots) = \sum\limits_a g_a(x,P|x_a)
\end{equation}
and, in accordance with the definition (4.8),
\begin{equation}\label{eq4.14}
\average{g(x,P)} = 0.
\end{equation}
If we substitute \eqref{eq4.13} into \eqref{eq4.12} and averaging the equation
obtained, taking into account \eqref{eq4.14}, we get
\begin{equation}\label{eq4.15}
P^\alpha\frac{\partial f}{\partial
x^\alpha}+\biggl(1+\frac{1}{2}\average{\xi^2}\biggr)
P^4\frac{\partial f}{\partial \eta} +P^4\average{\xi\frac{\partial
g }{\partial\eta}}-P_4P^4\frac{\partial}{\partial
P_\alpha}\average{(\partial_\alpha\xi)g}=0.
\end{equation}
Averaging \eqref{eq4.12} over all the particles except the ath one, taking
into account the results of Equation \eqref{eq4.15} in the equation
obtained, and suppressing terms quadratic with respect to $\xi$,
we shall find an equation to determine the correlation function
\begin{equation}\label{eq4.16}
g_a(x,P|x_a)\equiv g_a,
\end{equation}
\begin{equation}\label{eq4.16a}
P^i\frac{\partial g_a}{\partial x^i}=%
-\xi_a P^4\frac{\partial f}{\partial \eta}+%
P_4 P^4\frac{\partial f}{\partial P_\alpha}\partial_\alpha \xi_a.
\end{equation}
The   integrals   of   Equation   \eqref{eq4.16a}   are   the   integrals
of geodesic lines in a Robertson-Walker universe: namely,
\begin{equation}\label{eq4.17}
P_\alpha=\mbox{Const};\quad
x^\alpha=x^\alpha_0+\pi^\alpha(\eta-\eta_0),
\end{equation}
where $\pi^\alpha=P^\alpha/P^4=-P_\alpha/P_4=\mbox{Const}$. If we
integrate \eqref{eq4.16} along the trajectories of \eqref{eq4.17}, we find that
\begin{eqnarray}\label{eq4.18}
g_a=-\int\limits_{\eta_0}^{\eta}d\eta'
\xi_a\bigl(\bigl|\ \mathbf{r}'(\eta')-
\mathbf{r}_a\bigr|\bigr)\frac{\partial}{\partial\eta'}f[\mathbf{r}'(\eta),\eta',P_i]+\hspace{4cm}\nonumber\\
+P_4\int\limits_{\eta_0}^{\eta}d\eta'\partial_\alpha\xi_a(|\mathbf{r}'(\eta')-\mathbf{r}_a|)
\frac{\partial}{\partial P_\alpha}f[\mathbf{r}'(\eta),\eta',P_i]+
\dot{g}_a(x^\alpha-\pi^\alpha(\eta-\eta_0),P_i).
\end{eqnarray}
We shall expand in the integrands of \eqref{eq4.18} the function
$f[\mathbf{r}'(\eta'),\eta',P_i]$
\[f[\mathbf{r}'(\eta'),\eta',P_i] =f(\mathbf{r},\eta,P_i)-
(\eta-\eta')\frac{P_i}{P_4}\frac{\partial f}{\partial x^i}.\]
But according to \eqref{eq4.15} $P^i(\partial f/\partial x^i)=O(\xi^2)$;
therefore, the distribution function to within an accuracy here
required, can be factored outside the integral in (4.18). For the
same reason the function $g_a$ can be represented in the form of
an arbitrary linear operator acting on $f(x, P$ selecting it so as
to satisfy the condition of (4.14). Having noted this, we shall
proceed to calculate the integrals in \eqref{eq4.18}. A non-zero
contribution to these integrals is only provided by the regions
\begin{equation}\label{eq4.19}
\bigl|\ \mathbf{r}'(\eta')-\mathbf{r}_a\bigr|\equiv %
\bigl|\ \mathbf{r}-\mathbf{r}_a +\vec{\pi} (\eta-\eta')\bigr|\leq
r_0.
\end{equation}
A non-zero contribution to the averages in Equation \eqref{eq4.15} is
provided by the regions
\begin{equation}\label{eq4.20}
\bigl|\ \mathbf{r}-\mathbf{r}_a\bigr|\leq r_0.
\end{equation}
Therefore, we are interested in the values of correlation
functions at the intersection of the regions \eqref{eq4.19} and \eqref{eq4.20}.

To simplify the calculations we can put $\mathbf{r}_a = 0$ and
direct the velocity of the particle along the z-axis. The value of
the correlation function obtained by integration is, generally
speaking, time-depended; however, in the region \eqref{eq4.21}
\begin{equation}\label{eq4.21}
\eta-\eta_0\geq 2r_0,
\end{equation}
\begin{eqnarray}\label{eq4.22}
Q\equiv \int\limits_{\eta_0}^\eta \xi
d\eta'=2\mu\left[\ln\frac{r_0+\sqrt{r_0^2-\varrho^2}}{r_a+z_a}
+z_a\left(\frac{9}{5r_0} -\frac{\varrho^2}{2r_0^3}\right)-\right.\hspace{3.5cm}\nonumber\\
\hspace{3cm}-\frac{z_a^3}{6r^3_0}-\left.\left(\frac{9}{5r_0}
-\frac{\varrho^2}{2r_0^3}\right)\sqrt{r_0^2-\varrho^2}+\frac{1}{6r^3_0}(r_0^2-\varrho^2)^{3/2}\right],\\
\Psi_a\equiv \int\limits_{\eta_0}^\eta \partial_\alpha\xi
d\eta'=\xi\pi^\alpha + 2\mu
x^\beta(\delta^{\alpha\beta}-\pi^\alpha\pi^\beta)\times\hspace{5.2cm}\nonumber\\
\hspace{3cm}\times\left[-\frac{z_a}{\varrho^2
r_a}+\frac{\sqrt{r_0^2-\varrho^2}}{\varrho^2r_0}
-\frac{1}{r^3_0}(z_a-\sqrt{r_0^2-\varrho^2})\right].\nonumber
\end{eqnarray}
It is necessary to generalize the expressions obtained, in the
manner
\[z_a\to (\vec{\pi},\mathbf{r}-\mathbf{r}_a);\quad \varrho^2=|\mathbf{r}-
\mathbf{r}_a|^2-(\vec{\pi},\mathbf{r}-\mathbf{r}_a)^2;   \]
\[z_a\to |\mathbf{r}-\mathbf{r}_a|;\quad  x^\beta_a \to x^\beta-x^\beta_a. \]
Since $\average{\Psi_a}=0$, the correlation function can be
represented as
\begin{equation}\label{eq4.23}
g_a=-\frac{\partial
f}{\partial\eta}(Q-\average{Q})+P_4\frac{\partial f}{\partial
P_\alpha}\Psi_\alpha.
\end{equation}
Calculating the averages in accordance with \eqref{eq4.23}, we get
\begin{equation}\label{eq4.24}
\begin{array}{l}
{\displaystyle \average{\xi(Q-\average{Q})}=\average{\xi Q}=\frac{3}{2}r_0 N\left(\frac{2\mu}{r_0}
\right)^2K};\\[12pt]
{\displaystyle \average{\xi
\Psi_\alpha}=-\average{(\partial_\alpha\xi)Q}=
\pi^\alpha \xi^2};\\[12pt]
{\displaystyle
\average{(\partial_\alpha\xi)\Psi_\beta}=\frac{3}{2}\frac{N}{r_0}\left(\frac{2\mu}{r_0}
\right)^2(\delta_{\alpha\beta}-\pi_\alpha \pi_\beta)\Lambda},
\end{array}
\end{equation}
where
\[K=\frac{17}{160}-\frac{7}{144}\ln 2-\frac{29}{50}-
\frac{31}{240}+\frac{13}{180} \left(
\frac{19}{25}-\frac{5}{21}\right)\approx 0.11.\]
The last value of \eqref{eq4.24} is logarithmically divergent at the lower
limit $r\to 0$, which is caused by the coincident illegitimacy of
a linear approximation of the Einstein equations and a Born
approximation. According to \eqref{eq3.8} at $r\sim 2\mu$, the local
fluctuations of the metric become large, and at the same time the
deflection angle of massless particle becomes large; therefore,
integration in \eqref{eq4.24} is
\[\Lambda=\ln\frac{r_0}{2\mu}-\ln2+\frac{1}{3}.\]
Substitute \eqref{eq4.23}, with the acount of \eqref{eq4.24} into Equation \eqref{eq4.15}
\begin{eqnarray}
P^\alpha\frac{\partial f}{\partial
x^\alpha}+\biggl(1+\frac{1}{2}\average{\xi^2}\biggr)
P^4\frac{\partial f}{\partial \eta}-\frac{3}{2}r_0 N
\left(\frac{2\mu}{r_0} \right)^2 KP^4\frac{\partial^2f}{\partial
\eta^2} - \hspace{3cm}\nonumber\\
-\average{\xi^2}P^4P_\alpha\frac{\partial}{\partial\eta}\frac{\partial
f}{\partial
P_\alpha}+\average{\xi^2}P_4P^4\frac{\partial}{\partial
P_\alpha}\left(\frac{P_\alpha}{P_4}\frac{\partial f}{\partial\eta}
\right)-\hspace{3cm}
\nonumber\\
-\frac{3}{2}\frac{N}{r_0}\left(\frac{2\mu}{r_0} \right)^2 \Lambda
P_4P^4 \frac{\partial}{\partial
P_\alpha}\left[P_4\left(\delta_{\alpha\beta}-\frac{P_\alpha
P_\beta}{P^2_4} \right)\frac{\partial f}{\partial
P_\beta}\right]=0.\nonumber
\end{eqnarray}
If we change the order of differentiation in the fourth and fifth
term, respectively, suppressing the small term which is
proportional to $r_0/\eta$ and transposing the final term to the
right-hand side, we eventually obtain kinetic equation for
massless particles in the region \eqref{eq4.21}
\begin{equation}\label{eq4.25}
P^4\biggl(1+\frac{5}{2}\average{\xi^2}\biggr) \frac{\partial
f}{\partial \eta}+P^\alpha\frac{\partial f}{\partial
x^\alpha}=\frac{3}{2}\frac{N}{r_0}\left(\frac{2\mu}{r_0} \right)^2
\Lambda P_4P^4 \frac{\partial}{\partial
P_\alpha}\left[P_4\left(\delta_{\alpha\beta}-\frac{P_\alpha
P_\beta}{P^2_4} \right)\frac{\partial f}{\partial P_\beta}\right].
\end{equation}
The term on the right-hand side of Equation \eqref{eq4.25} is a collision
'integral', it describes the process of altering the momentum of
massless particles by interaction with the local gravitational
fields. The supplementary term on the left-hand side of \eqref{eq4.25},
$\frac{5}{2}\average{\xi^2}P_4\partial f/\partial \eta$ describes
the effective change in the velocity of a particle in local
gravitational fields. Indeed, if we suppress the collision term in
\eqref{eq4.25}, then the equations of characteristics will be of the form
\[P_\alpha=\mbox{Const};\quad  x^\alpha=
\frac{\pi^\alpha(\eta-\eta_0)}{1+\frac{5}{2}\average{\xi^2}} +x^\alpha_0. \]
These equations describe the motion of a particle with the
velocity
\begin{equation}\label{eq4.26}
v=c\biggl(1-\frac{5}{2}\average{\xi^2}\biggr)<c.
\end{equation}
A local observer who measures the velocities of massless particles
equal to the speed of light and assumes that medium velocity of
these particles coincides with the speed of light will observe a
local violation of the law of conservation of the particle number.
Indeed, if the observer knows that a massless particle escapes
from a certain point in space in his direction he has every reason
to expect to register it within a completely definite time $\Delta
t= \Delta l/c$. If this fails to occur, the observer will record a
local violation of the law of conservation of the particle number.
However, since all massless particles travel at identical
velocities (4.26), the observer can define this velocity as the
speed of light. In agreement with this operation is the
renormahzation of the momentum in \eqref{eq4.25}, in which the term on the
left-hand side of the equation, quadratic with respect to $\xi$,
vanishes.

\section{The Laws of Conservation and Generalized Kinetic
Equations}
Carrying out the foregoing renormahzation of the momentum and
changing over from a synchronous frame of reference, to an
arbitrary one, we shall write Equation \eqref{eq4.25} in an invariant form
\begin{equation}\label{eq5.1}
[H,f]=-\frac{\partial I_i}{\partial P_i},
\end{equation}
where
\begin{equation}\label{eq5.2}
I_i=\frac{3}{2}\frac{N}{r_0a}\left(\frac{2\mu}{r_0}
\right)^2\Lambda W_{ik}\frac{\partial f}{\partial P_k},
\end{equation}
\begin{equation}\label{eq5.3}
W_{ik}=(u,P)[P_iP_k+g_{ik}(u,P)^2-(u,P)(u_i P_k-u_kP_i)].
\end{equation}
The coefficient in \eqref{eq5.2} can be given a clear physical meaning by
taking account of the determination of the particle mass ($m = \mu
a$) and the particle number $N$. Let us introduce the medium mass
density $\varrho_m=Nm/4/3\pi(ar_0)^3$, which corresponds to
massive particles. It should be noted that this value changes as
$\varrho_m \sim t^{-4/3}$ at the linear stage of accretion as
distinct from the total mass density $\varrho\sim t^{-2}$. Then
\eqref{eq5.2} takes the form
\begin{equation}\label{eq5.2a}
I_i=8\pi\Lambda\varrho_m W_{ik}\frac{\partial f}{\partial P_k}.
\end{equation}
The symmetric tensor $W_{ik}$ has the following properties:
\begin{equation}\label{eq5.4}
W_{ik}P^k=0;\quad W_{ik}u^k=0.
\end{equation}
Let us integrate \eqref{eq5.1} over the space of momenta
\[n^i_{,i}=0, \]
where
\[n^i=\int P^i f\ \delta(H)d^4P\]
is the density vector of the particle-number flux. Thus the local
law of the conservation of particle number is also satisfied for a
non-renormalized equation \eqref{eq4.25}, which can be verified by
integrating \eqref{eq4.25} over the space of momenta and the entire
three-dimensional space. We shall multiply \eqref{eq5.1} by $P^i$ and
integrate over the space of momenta. If we integrate twice by
parts on the right-hand side of the equation, we get, taking into
account \eqref{eq5.4}
\begin{equation}\label{eq5.5}
T^{ik}_{,k}=32\Lambda\varrho_m m(g^{ik}-u^iu^k)T_{kl}u^l,
\end{equation}
where
\[T^{ik}=\int P^i P^k f\ \delta(H)d^4P \]
is the energy-momentum tensor of massless particles. The energy of
massless particles is conserved, which can be verified by
comparing \eqref{eq5.5} with the time-like vector of conformal motion
$xi^i=\delta^i_4$:
\[{\cal P}^i_{,i}=0,  \]
where ${\cal P}^i=T^{ik}\xi_k$. However, the three-dimensional
momentum of the particle is not conserved, which is the
consequence of the non-conservative nature of the system of
massless particles, - a part of the momentum of massless particles
is transferred to massive ones.

The kinetic equation \eqref{eq5.1} admits a natural generalization in the
case where massive particles travel at arbitrary, but small
velocities at which the rate of accretion is not altered
significantly. Let $F(\eta, P')$ be a function of the distribution
of massive particles with masses which are now arbitrary, so that
\[n^i_m=\int F(\eta,P')P'^i d^4P' \]
is the density vector of the flux of massive particles. Then it is
known that collision term of \eqref{eq5.2} ($P'_i=mu_i$) can be written as
\begin{equation}\label{eq5.6}
I_i=8\pi\int\Lambda m^2 F(P')W_{ik}\frac{\partial f}{\partial
P_k}d^4P'.
\end{equation}
To make sure of this it is sufficient to put $F(P') =
m^{-1}\delta(\mathbf{P})\delta(H_m/2)$, i.e., to consider the
distribution of fixed-mass particles at rest. The collision term
of \eqref{eq5.6} takes account of the processes of transferring a momentum
from massless particles to massive ones, but fails to take account
of the reverse processes. To take into account the latter it is
necessary to add to \eqref{eq5.6} a term which is antisymmetric respect to
the transposition of particles
\begin{equation}\label{eq5.7}
I_i=8\pi\int d^4P'\Lambda m^2 W_{ik}\left[F(P')\frac{\partial
f(P)}{\partial P_k}-f(P')\frac{\partial F(P')}{\partial
P'_k}\right].
\end{equation}
The kernel $W_{ik}$ of the collision integral thus obtained is the
same as that of the Belayev-Budker (1956) collision integral, if
in the latter we let the momentum of one of the particles tend to
infinity. The difference lies in the mutiplier $8\pi$ (instead of
$2\pi$ in Belayev and Budker (1956)), which is the consequence of
the well-known effect, viz., the deflection angle of a photon in a
gravitational field with the scope of Einstein's theory is two
times as great as its Newtonian value.

The collision integral of \eqref{eq5.7} as distinct from \eqref{eq5.1} now
satisfies all the necessary laws of conservation. It can be used
as the basis for showing that the total of energy and momentum
within the system of 'dust + massive particles + massless
particles' is conserved.

\section{Isotropization of Homogeneous Distribution of Massless
Particles}
An exact solution of the kinetic equation \eqref{eq5.1} (as well as
\eqref{eq4.25}) is an arbitrary isotropic distribution $f(P)$. If the
initial distribution is anisotropic, however, it will be
isotropized by gravitational interactions. Let the distribution of
massless particles at the momentum of time $\eta=\eta_1$ in a
synchronous frame of reference be of the form
\begin{equation}\label{eq6.1}
f(\eta_1,P_\alpha)=\sum\limits_{l,m}f_{lm}(\eta_1,P)\mbox{Y}^l_m(\theta,\varphi),
\end{equation}
where $\theta$ and $\varphi$ are the azimuthal and the polar
angles in the momentum space, respectively. Representing the
distribution of $f(\eta,P_\alpha)$) in a form analogous to \eqref{eq6.1},
and on separating the variables, we obtain equations for the
functions $f_{lm}(\eta,P)$,
\[\frac{\partial f_{lm}}{\partial\eta}=-l(l+1)m\Lambda a f_{lm} \varrho_m 8\pi,\]
\begin{equation}\label{eq6.2}
f_{lm}(\eta,P)=f_{lm}(\eta_1,P)
\exp\left[-l(l+1)8\pi\int\limits_{t_1}^t\Lambda m \varrho_m
dt'\right].
\end{equation}
In particular, for the dependence of $m(t)$ and $\varrho_m(t)$ we
have obtained, we shall find by integrating in (6.2):
\begin{equation}\label{eq6.3}
f_{lm}(\eta,P)=f_{lm}(\eta_1,P) \exp\left[-l(l+1)24\pi\Lambda m
\varrho_m \left( 1-\left(\frac{t_1}{t}\right)^{1/3}
\right)\right],
\end{equation}
where $m = m(t)$, $\varrho_m = \varrho_m(t)$. At $t\to\infty$ the
expression in the exponent increases proportional to $t^{1/3}$.
Therefore, at $t\to\infty$ the expression (6.1) retains only one
harmonic with $l= 0$, i.e., the distribution is isotropized. The
$l$-harmonic is presented by the angular scale
$\Delta\varphi=2\pi(l +1)$. Therefore, at the given present-time
values the particle masses $m(t)$ and their medium density
$\varrho_m(t)$ all harmonics with the angular dimension
\begin{equation}\label{eq6.4}
\Delta\varphi2\pi<\sqrt{24\pi\Lambda m\varrho_m t}
\end{equation}
vanish. At $m \sim 10^{16}M_\odot$, $\varrho\sim 10^{30}\mbox{g
cm}^{-3}$ and $t = 2\times10^{10}$ yr we obtain from \eqref{eq6.4}
$\Delta\varphi < 10$ angular minutes (Figures 1 and 2).
\Fig{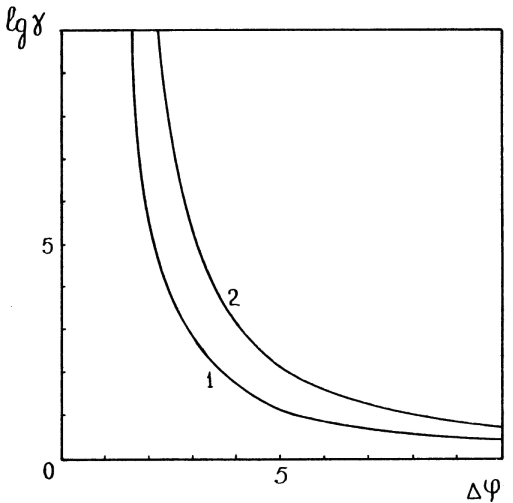}{The dependence of the damping coefficient of the
angular harmonic distribution $\gamma=f_{lm}(t_0)/f_{lm}(t)$ from
the angular scale $\Delta\varphi=2\pi(l +1)$. Curve  1: $m=
10^{16}M_\odot$, $\varrho_m/\varrho=0.1$. Curve 2: $m=
10^{16}M_\odot$, $\varrho_m/\varrho= 0.2$. The angle
$\Delta\varphi$ is measured in angular minutes.}

Thus, any relict radiation must be highly homogeneous on scales
less than 10 angular minutes. This effect can resolve the
contradiction between the deductions of a standard adiabatic
theory of galaxy formation (as in Zeldovich, 1983) and observation
data which give evidence in favour of the absence of small-scale
fluctuations of relict radiation.

Now let us assume that the accretion is completed at the moment of
time $t_l$ i.e., $m = \mbox{Const}$, and $\varrho_m\sim t^{-2}$.
Then in place of (6.3) we obtain from \eqref{eq6.2},
\begin{equation}\label{eq6.5}
f_{lm}(\eta,P)=f_{lm}(\eta_1,P) \exp\left[-l(l+1)8\pi\Lambda m
\varrho_m \frac{t^2}{t_1}\left( 1-\frac{t_1}{t}\right)\right].
\end{equation}
In case, at $t\sim\infty$ the exponential index becomes large,
$4\Lambda l(l + l)m/3t_l$. This implies that harmonics with an
angular scale of the order of tens of degrees can be dampened, and
isotropization takes place according to \eqref{eq6.5} at the earliest
stages following the termination of accretion. Apparently, this
cannot be the case. But it follows then that the masses of super
aggregations should be either less than $10^{16} M_\odot$, or
their formation should be completed at significantly later moments
of time $t_1 > 10^{16}$ s.

\Fig{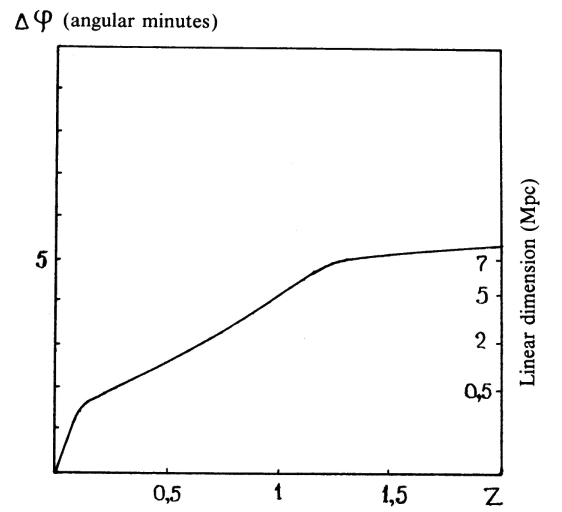}{The dependence of the maximum resolution of
$\Delta\varphi$ on the red shift $Z$: $m = 10^{15}M_\odot$;
$\varrho_m/\varrho= 1$. The graph shows the values of
$\Delta\varphi$ for which $\gamma=e$.}

\section*{Acknowledgements} In conclusion, Yu. G. Ignat'ev thanks G.
S. Bisnovatyi-Kogan and I. D. Novikov for a highly fruitful
discussion of the paper. The authors also thank the participants
of the seminar on general-relativity statistics and cosmology of
the Kazan Teachers Training Institute for a helpful discussion of
the work.

\begin{comment}
\vskip 24pt\noindent \refstepcounter{figure}%\setcounter{figure}{1}
\epsfig{file=12.eps,width=8cm}\label{ris12} \vskip 12pt \noindent
{Figure \bf \thefigure.}\hskip 12pt{\sl Influence of the second
GMSW parameter $\Upsilon$ on the GW energy absorbtion
$\Delta\varepsilon_g/\varepsilon_g=(\beta^2_0-\beta^2)/\beta^2_0$
by $\xi^2=0.0045$, $\gamma_\perp=1/6$: $\Upsilon=3$  (solid line),
$\Upsilon=10$ (dashed line), $\Upsilon=100$ (dotted line). \hfill}
\vskip 12pt\noindent%
\end{comment}
%

\end{document}